\documentclass[aps,twocolumn,showpacs]{revtex4}

\usepackage{epsfig}

\begin{document}

\title{Optical implementation of Deutsch-Jozsa and Bernstein-Vazirani\\
quantum algorithms in eight dimensions}

\author{E. Brainis,$^1$ L.-P. Lamoureux,$^2$ N. J. Cerf,$^2$ 
Ph. Emplit,$^1$ M. Haelterman,$^1$ and S. Massar$^{3,2}$}
\affiliation{$^1$ Optique et Acoustique, CP 194/5, Universit\'e Libre de
Bruxelles, Av. F. D. Roosevelt 50, 1050 Bruxelles, Belgium\\
$^2$ Th\'eorie de l'Information et des Communications, CP 165/59, 
Universit\'e Libre de Bruxelles, Av. F. D. Roosevelt 50, 
1050 Bruxelles, Belgium\\
$^3$ Service de Physique Th\'eorique, CP 225, Universit\'e Libre de
Bruxelles, Bvd. du Triomphe, 1050 Bruxelles, Belgium}

\begin{abstract}
We report on a fiber-optics implementation of the Deutsch-Jozsa and
Bernstein-Vazirani quantum algorithms for 8-point functions. The measured 
visibility of the 8-path interferometer is about 97.5\%. Potential
applications of our setup to quantum communication or cryptographic
protocols using several qubits are discussed.
\end{abstract}

\pacs{03.67.Lx, 42.50.-p}
\maketitle

The last decade has seen the emergence of the field of quantum
information processing. A particularly  promising application is the
concept of quantum algorithms, which allow certain problems
such as factorization\cite{Shor} or searching\cite{Grover}
to be solved much faster than on a classical computer.
Another algorithm which we will be interested in here 
is Deutsch's algorithm\cite{Deutsch},
the first quantum algorithm ever discovered, which was later
generalized by Deutsch and Jozsa\cite{DeutschJozsa} (DJ). 
The DJ algorithm discriminates between a constant or a balanced
$N$-point binary function using one single quantum query, while a classical
algorithm requires ${\cal O}(N)$ classical queries.
It was later on adapted by Bernstein and Vazirani (BV) for efficiently
querying a quantum database\cite{BV,BS}.

In the present paper, we report on a fiber-optics implementation of the DJ
algorithm using standard telecom optical components and a single-photon 
detector. The DJ algorithm has already been implemented using NMR\cite{nmr} 
(see also \cite{nmr2} for a NMR implementation of the
BV algorithm), table-top optics\cite{DJoptical} (optical demonstrations 
of other quantum algorithms also include Grover's 
algorithm\cite{Kwiat,Amsterdam}), molecular states\cite{molecular}, 
and very recently using an ion trap\cite{trap}. However, our setup
separates from these realizations (especially that of \cite{DJoptical})
on several major aspects. First, it relies on guided optics components,
which makes it unnecessary to perform a precise alignment, 
and it is made robust against phase fluctuations 
by use of an autocompensation technique.
Second, although it relies on linear optics,
our realization is relatively efficient in terms of used optical resources
compared to standard linear optical implementations of quantum computation.
The central idea of such implementations consists 
in representing the basis states of a $N$-dimensional Hilbert space 
by $N$ optical paths so that unitary transformations are obtained by chaining
linear optics components that make these paths 
interfere\cite{Reck,CerfKwiat}. Such implementations seem, however, to be
inherently inefficient since the space requirement (the number of optical
components) and the time requirement both
grow exponentially with the number $n$ of qubits (with $N=2^n$) \cite{fn}.
In contrast, in our setup, the number of components 
is kept linear in $n$, while the time needed still grows exponentially.
Note that any implementation of an algorithm involving an
arbitrary $2^n$-point function (also called oracle)
does in any case require exponential resources to simulate this function.
Therefore, the linear optical implementation of quantum algorithms
involving oracles can reasonably be made as efficient as 
any other implementation in this respect.
For all these reasons,
our experimental demonstration works with a 8-point (3-qubit) 
function and might probably be extended even further 
without fundamental difficulty, 
while today's largest size optical demonstrator of the DJ algorithm
involves a 4-point function\cite{DJoptical}.

\begin{figure*}
\centerline{\includegraphics[width=17.5truecm]{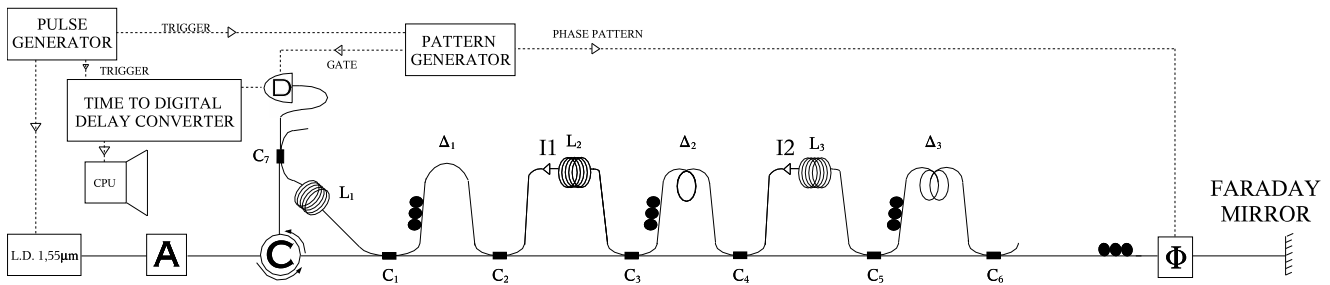}}
\caption{Fiber optics setup implementing the 8-dimensional Deutsch-Jozsa and
Bernstein-Vazirani quantum algorithms.}\label{fig:gfx}
\end{figure*}

Let us start by recalling the principle of the DJ algorithm. 
At the core of the algorithm is the oracle
which computes a function $f(x)$, where $x\in\{0,1\}^n$
is an $n$ bit string, and $f\in\{0,1\}$ is a single bit. The
DJ problem is to discriminate whether $f$ is a constant or balanced
function, while querying the oracle as few times as possible. A balanced
function is such that the number of $x$'s on
which $f(x)=0$ is equal to the number of $x$'s on which $f(x)=1$.
Classically, $2^{n-1}+1$ queries are necessary in the worst
case, whereas the DJ algorithm requires a single query as we shall see.
In this algorithm, $n$ qubits are used, and the basis of 
the Hilbert space is chosen as
$|x\rangle = |x_1 x_2 \ldots x_n\rangle$ where $x_i \in \{0,1\}$.
The quantum oracle carries out the transformation
\begin{equation}
|x\rangle |y\rangle  \stackrel{\rm oracle}\longrightarrow  |x\rangle
|y \oplus f(x)\rangle \ ,
\label{oracle1}
\end{equation}
where $|y\rangle$ is an ancilla qubit.
By choosing $|y\rangle=(|0\rangle - |1\rangle)/\sqrt{2}$, 
the action of the oracle simplifies into
\begin{equation}
|x\rangle \stackrel{\rm oracle}\longrightarrow  (-1)^{f(x)}|x\rangle \ .
\label{phaseoracle}
\end{equation}
since $|y\rangle$ then remains unchanged.


The DJ algorithm starts with the system in the state
$|0\rangle=|00 \ldots 0\rangle$. Next, a Hadamard transform $H$ is
applied independently on each of the $n$ qubit.
Using the definition $H |x\rangle = 2^{-n/2} \sum_{z\in
\{0,1\}^n} (-1)^{x \cdot z} |z\rangle $, where $x\cdot z =
\sum_i x_i z_i \bmod 2$ is the inner product of two $n$-bit strings,
we see that the Hadamard transform acting on the initial state simply
yields a uniform superposition of all states: $H |0\rangle =
2^{-n/2}\sum_{x} |x\rangle $. This state is then
sent through the oracle whereupon it becomes 
\begin{equation} \label{oracle}
2^{-n/2}\sum_x (-1)^{f(x)} |x\rangle \, . 
\end{equation}
The superposition principle
allows the oracle to be queried on all input values
in parallel. A second Hada\-mard transform is then carried out to
obtain the state
\begin{equation}
 2^{-n} \sum_{x,z}  (-1)^{x \cdot z + f(x)} |z\rangle \label{DJwavefunction}
\end{equation}
which is finally measured in the $z$ basis.
One easily deduces from Eq. (\ref{DJwavefunction}) that
when $f$ is constant, the probability of measuring $|0\rangle$ is one.
In contrast, when $f$ is balanced, 
this probability is always zero, so the
DJ algorithm can distinguish with certainty
between these two classes of functions by querying the oracle a single time.
The BV variant of this algorithm is also based on the transformation
leading to Eq. (\ref{DJwavefunction}). Suppose that the oracle is
restricted to be of the form
$f_j(x) = x \cdot j$
where $j\in \{ 0,1\}^n$ is an arbitrary $n$-bit string. 
The aim is to find the bit string $j$ with as few queries
as possible. Classically one needs at least $n$ queries since each query
provides one independent bit of information at most about $j$. 
Quantum mechanically, a single query
suffices since using $f_j(x)$ in
Eq. (\ref{DJwavefunction}) shows that the measurement outcome is
$z=j$ with  probability.



Our all optical fiber (standard SMF-28) setup is illustrated 
in Fig.~1. Initially, a 3~ns light pulse is
produced by a laser diode at $1,55~\mu$m,
attenuated by an optical attenuator (Agilent 8156A), and then
is processed through three unbalanced Mach-Zehnder (M-Z) 
interferometers with path length differences $\Delta_l$ ($l=1,2,3$)
obeying $\Delta_3 = 2 \Delta_2 = 4 \Delta_1 = 15$~ns.
Each MZ interferometer doubles the number of pulses so that, 
at the coupler C6, we get eight equally spaced pulses. 
This corresponds to the action of the Hadamard transform on the three input
qubits in state $|0\rangle$. 
The pulses are then reflected by a Faraday mirror 
and, on their way back, are modulated by a phase modulator (Trilink) 
commanded by a pattern generator (Agilent 81110A)
which selectively puts a phase shift of $0$ or $\pi$ on each pulse 
according to the 8-point function [see Eq. (\ref{oracle})]. 
The pulses then pass back through the three M-Z interferometers, 
thereby realizing a second triple Hadamard transform,
and are sent via a circulator
to a single-photon detector (id Quantique id200),
which completes the DJ or BV algorithm.
The additional delay lines $L_1$, $L_2$, and $L_3$,
which obey $L_3 > 2 L_2 >  4 L_1 > 8 \Delta_3$, 
ensure that the different outputs of the DJ or BV algorithm all
reach the detector at different times.
The delay lines $L_2$ and $L_3$ also contain an isolator
so that the pulses are not transmitted on their way 
to the mirror. The photodetector was gated during 5~ns around the arrival time 
of each pulse by the pattern generator.
The output of the detector was registered 
using a time to digital delay converter (ACAM-GP1) connected to a computer.
All delays $\Delta_i$ and $L_i$ were chosen to be integer multiples 
of $\Delta_1$ within 0,2~ns. All electronic components were
triggered by a pulse generator (Standford Research Inc. DG535).  
In order to maximize the visibility, polarization controllers were introduced 
in the long arm of each M-Z interferometer and in front of the
polarization-sensitive phase modulator.
Once optimized, the setup was stable for days.


This implementation of the DJ and BV algorithms
differs from an earlier optical implementation of the DJ
algorithm\cite{DJoptical} in several important aspects.
First, we run the algorithm for $n=3$ qubits and, more importantly, 
we measure all 8 outcomes, which makes it possible to realize 
the BV algorithm as well (the previous implementation \cite{DJoptical} 
works with $n=2$ qubits and only measures the outcome $z=0$). 
Second,  we operate at telecom wavelengths 
in optical fibers using a setup closely inspired from 
the ``plug-and-play'' quantum cryptographic system developed by Gisin and
collaborators (see e.g. \cite{plugandplay}). For this reason, the present
setup in a slightly modified version can be adapted to implement
quantum cryptography using higher dimensional systems\cite{highdQcrypto}
or to illustrate quantum communication complexity protocols\cite{BCW}
over distances of a few kilometers. These potential applications 
will be discussed below. Third, the used resources are quite
different in the two implementations: 
when scaled to a large $n$, the implementation of \cite{DJoptical}
requires exponential time and exponentially many optical elements, whereas
our implementation also requires exponential time but only 
a linear number of optical elements. This is because
the $n$ qubits are realized as $2^n$ separate optical paths 
in \cite{DJoptical}, whereas in our case they are represented as $2^n$
light pulses traveling in a single optical fiber, extending naturally
the ``time-bin'' realization of qubits used in \cite{plugandplay}. 
The small number of
optical elements in our setup therefore implies that it is relatively easy to
increase the number of qubits $n$ while keeping the optical setup stable.
As we shall see, a disadvantage of our setup is that the Hadamard 
transform can only be implemented with a probability of success of $1/2$.
Since $2n$ Hadamard transforms are needed for the DJ algorithm with 
$n$ qubits, the resulting attenuation is $2^{-2n}$.

Let us now prove that our optical setup indeed realizes
the DJ and BV algorithm.
The quantum state describing the eight pulses 
at coupler C6 can be written as
\begin{eqnarray}
|\psi\rangle \propto
\sum_{x=000}^{111}
\exp\left[ i \sum_{l=1}^3
(k x_l \Delta_l + \pi x_l ) \right] |\sum_{l=1}^3 x_l \Delta_l\rangle
\end{eqnarray}
where $x$ stands for $(x_1,x_2,x_3)$ and
$|p\rangle$ denotes a pulse located at position $p$.
The three bits $x_1,x_2,x_3=0,1$ label whether the pulse took 
the short ($x=0$) or the long ($x=1$) path through each interferometer.
The factor $\exp[i k \sum_l x_l \Delta_l]$ with $k$ being the wave number
takes into account the phase difference between a pulse traveling
along the short or long paths of the interferometers. The factor
$\exp[i \sum_l \pi x_l ]$ takes into account the phase
accumulated at the couplers: if the pulse takes the short path, it
is transmitted at two couplers, whereas, if it
takes the long path it is reflected twice.
After reflection at the Faraday mirror and phase modulation, 
the pulses cross again the three M-Z interferometers
and reach the photodetector in the state
\begin{widetext}
\begin{eqnarray}
|\psi\rangle \propto \sum_{x,y, z}
(-1)^{f(x)} 
\exp\left[ i  \sum_{l=1}^3 	\left(
 k  (x_l + y_l) \Delta_l
+ \pi ( x_l + y_l - y_l (z_l+z_{l+1}) + {z_l+z_{l+1}\over 2}) 
\right)
\right] 
|  \sum_{l=1}^3 ((x_l + y_l) \Delta_l + z_l L_l) \rangle
\end{eqnarray}
where the bits $y_1,y_2,y_3$ equal $0$ or $1$ according to whether
the pulse passed through the short or the
long path of each of the interferometers on its way back, and the bits
$z_1,z_2,z_3$ equal $0$ or $1$ according to
whether or not the pulse exited each of the interferometer in the path 
containing the delay lines $L_1$, $L_2$ or $L_3$.
Note that we put $z_4=0$.
We have again taken into account the phases
induced by transmission or reflexion at the couplers C1-C6.
The final state contains 120 pulses, but 
we are only interested in the eight pulses
such that $x_1+y_1 = x_2+ y_2 = x_3 + y_3 = 1$, which are those 
that exhibit 8-path interference. The other pulses are filtered out
in the computer analysis (they correspond to different time bins).
The final state then becomes
\begin{eqnarray}
|\psi\rangle \propto
\sum_{z}
(-i)^{z_1} (-1)^{z_2 + z_3}
\sum_{x}
(-1)^{f(x_1,x_2,x_3)+
x_1 (z_1+z_2) + x_2 (z_2+z_3) + x_3 z_3}
| \Delta_1 + \Delta_2 +  \Delta_3
+ z_1 L_1 + z_2 L_2 + z_3 L_3
\rangle \label{final}
\end{eqnarray}
\end{widetext}
By relabeling the time bins according to the substitution
$z_1\to z_1+z_2+z_3$ mod~2, $z_2 \to z_2+z_3$ mod~2, and $z_3 \to z_3$, 
this equation coincides (up to irrelevant phases and 
an overall normalization factor)
with Eq.~(\ref{DJwavefunction}) with the 8 logical states
$|z\rangle$ identified as specific time bins.


The setup thus realizes the DJ or BV algorithm, 
the main difference with an ideal algorithm being an
extra attenuation by a factor of $2^{-7}$. A factor $2^{-3}$
originates from the couplers C2, C4, and C6 
because each time a pulse passes through
these couplers it only has a probability $1/2$ of exiting
by the right path. Otherwise, it is absorbed
by the isolators I1 and I2 or by the unconnected fiber pigtail 
at coupler C6. Another factor $2^{-3}$ is due to
the filtering out of the 112 pulses produced on the way back
that do not correspond to 8-path interferences.
The remaining factor $2^{-1}$ is due to the coupler C7.
This overall loss of 21~dB could be remedied 
by replacing the couplers C2, C4, C6, and C7 by
optical switches which would direct the light pulses along the appropriate
path. High speed, low loss optical switches are not
available commercially at present, so we had to use couplers
in the present experiment. However, we emphasize that 
this is a technological rather than a fundamental limitation.

In order to characterize the performances of our setup,
we considered the $2^n$ oracles of the form $f_j(x)=x\cdot j$ and
${\overline f}_j(x)= x \cdot j + 1 {\rm ~mod~} 2$ (i.e., the oracles in
the BV algorithm and their complements). For each oracle $f_j$
(or ${\overline f}_j$), we ran the algorithm 500,000 times and registered
the number of counts in time bin $z$, denoted as
$N_j(z)$ (or ${\overline N}_j(z)$).
The algorithm gives constructive interference 
in the time bin $z=j$ for the oracle $f_j$ or ${\overline f}_j$,
and destructive interference elsewhere.
We then computed
\begin{equation}
V_j(z) = {1 \over 2} \left (
{N_z(z) - N_j(z) \over N_z(z) + N_j(z) }
+ {{\overline N}_z(z) - N_j(z) \over {\overline N}_z(z) + N_j(z) } 
\right) 
\end{equation}
for each pair of oracles with $j\ne z$, and calculated the
visibility $V(z)$ in time bin $z$ by 
taking the average of $V_j(z)$ over all values of $j$.
The measured visibilities $V(z)$ for 2 and 3 qubits are shown in Table~I.
Remarkably, they remain relatively high when going 
from 2 to 3 qubits in spite of the fact that
8 path interferences are involved. This is because the path differences are
automatically compensated and only $n+1$ polarizations must
be adjusted. It should therefore be relatively easy to go beyond $n=3$
without significantly decreasing the visibilities.

Because of the attenuation in our setup along with the quantum efficiency 
($\simeq 10.5$~\%) and the dark count probability ($\simeq 10^{-4}$~ns$^{-1}$)
of our detector, the signal-to-noise ratio was not high enough to perform
these visibility measurements in the single-photon regime. For this reason,
in our experiment, approximatively 20 photons per pulse entered 
the phase modulator (oracle) in 4 dimensions, and approximatively 50
in 8 dimensions. 
However, minimal modifications should allow
us to decrease the number of photons while keeping the
signal-to-noise ratio constant. In particular, by reducing the pulse length
or using two detectors instead of the coupler C7,
it should be possible to operate
in the single-photon regime in 4 (and possibly
8) dimensions. Moreover, as already mentioned,
the Hadamard transform could be rendered deterministic 
by using fast low-loss optical switches instead of couplers, 
which would strongly reduce the losses.



\begin{table}[t]
\begin{tabular}{lllllllll}
\hline
z & 1 & 2 & 3 & 4 & 5 & 6 & 7 & 8\\
\hline
$V(z)^{n=2}$ & 98.4 & 97.4 & 98.5 & 98.6 &&&&\\ 
$V(z)^{n=3}$ & 96.78 & 97.99 & 97.68 & 97.32 & 97.33 & 97.56 & 97.37 & 97.28\\
\hline
\end{tabular}
\caption{Measured average visibility $V(z)$ in the $z$th time bin 
for the DJ or BV algorithm with $n=2$ and $n=3$ qubits.} 
\end{table}

The present experiment can be extended in several ways. For example,
one could implement the distributed Deutsch-Jozsa problem\cite{BCW}
where two parties, which receive each as input a $2^n$-bit string 
(denoted as $f$ and $g$), 
must decide whether $f=g$ or $f$ differs from $g$ in exactly $2^{n-1}$
bits (they are promised that only one of these two cases can occur). 
The value of $n$ for which the gap
between the classical and quantum algorithms sets in is unknown, 
but recent results suggest that it could be as early as $n=4$ \cite{B}.
The distributed Deutsch-Jozsa algorithm 
could be realized with a slight modification 
of our setup in which 
the two parties, separated by by a few kilometers of optical fiber,
would use each a phase modulator.
This quantum protocol can also be easily adapted to multidimensional
quantum cryptography, where two parties randomly choose 
their patterns $f$ and $g$. By publicly revealing part of
$f$ and $g$, the parties can use the correlations between the measurement
outcomes to establish a secret key.  
A detailed analysis shows that for n = 2, this exactly coincides 
with the 4-dimensional cryptosystem based on 2 mutually
conjugate bases, which has been shown
to present advantages over quantum cryptography 
in two dimensions\cite{multidimcrypto}.
A final potential application of this setup is to test
quantum non locality using the entanglement-based
Deutsch-Jozsa correlations\cite{BCT}. 
An entangled state of $2^n$ time bins must be produced, for
example using the source \cite{RMZG}, and each party must then 
carry out phase modulation and a Hadamard transform
(which we have demonstrated are easy to realize on time bin entangled
photons). 
The correlations between the chosen phases and the measured time 
of arrival of the photons at each side
should exhibit quantum non locality. It has recently been
shown that these correlations are non local for $n=4$ \cite{B},
and that they exhibit exponentially strong
resistance to detector inefficiency for large $n$ \cite{M}, 
which means here that the 3~dB losses 
at each Hadamard transform can in principle be tolerated. 

In summary, we have demonstrated, by implementing the DJ and BV algorithms 
for 3 qubits, a simple and robust method for manipulating 
multidimensional quantum information encoded in time bins
in optical fibers. We anticipate that our method will have wide
applicability for quantum information processing and quantum
communication using higher dimensional systems.

It is a pleasure to thank N. Gisin, W. Tittel,
G. Ribordy, H. Zbinden, and all the members of {\it GAP optique} for very
helpful discussions and for providing the single-photon detector. 
We acknowledge financial support from the Communaut\'e Fran\c{c}aise 
de Belgique under ARC 00/05-251,
from the IUAP programme of the Belgian government under grant V-18, and
from the EU under project EQUIP (IST-1999-11053).



\begin{thebibliography}{99}
\bibitem{fn} This exponential growth can be overcome in some cases (e.g.
the optical implementation of Grover's algorithm of \cite{Amsterdam}, in which
the number of optical elements remains constant although the space and time
still grow exponentially).
\end{thebibliography}
\end{document}